\documentclass[conference]{IEEEtran}
\IEEEoverridecommandlockouts
\usepackage{cite}
\usepackage{amsmath,amssymb,amsfonts}
\usepackage{algorithmic}
\usepackage{graphicx}
\usepackage{textcomp}
\usepackage{xcolor}
\usepackage[T1]{fontenc} 
\def\BibTeX{{\rm B\kern-.05em{\sc i\kern-.025em b}\kern-.08em
    T\kern-.1667em\lower.7ex\hbox{E}\kern-.125emX}}
\usepackage{fancyhdr}

\begin{document}

\title{Comparing quaternary and binary multipliers }

\author{\IEEEauthorblockN{ Daniel Etiemble}
\IEEEauthorblockA{\textit{Computer Science Laboratory (LRI)} \\
\textit{Paris Saclay University}\\
Orsay, France \\
de@lri.fr}

}
\maketitle

\begin{abstract}
We compare the implementation of a 8x8 bit multiplier with two different implementations of a 4x4 quaternary digit multiplier. Interfacing this binary multiplier with quaternary to binary decoders and binary to quaternary encoders leads to a 4x4 multiplier that outperforms the best direct implementation of a 4x4 quaternary multiplier. The far greater complexity of the 1-digit multipliers and 1-digit adders used in this direct implementation compared to the binary 1-bit multipliers and full adders cannot be compensated by the reduced count of quaternary operators. As the best quaternary multiplier includes the corresponding binary one, it means that there is no opportunity to get less interconnects, less chip area, less power dissipation with the quaternary multiplier.
\end{abstract}

%\make

\section{Introduction}\label{sec1}
Since the 50's, many implementations of multivalued circuits have been proposed. In the last decade, most proposals used the CNTFET technology. 

Most presented implementations of ternary or quaternary circuits claim advantages of multiple valued circuits. The following quote summarizes the arguments  that may be found in most of these papers : 
``MVL circuits have potential advantages. Using MVL circuits reduces the complexity of interconnection via reducing the number
of wires since each wire carries more than one digit of data. Power consumption and area of the MVL circuits are generally less than the corresponding binary circuits due to the reduction in number of active elements \cite{Haixia}.

We examined ternary circuits in \cite{Eti1} and we compared the implementations of ternary adders and multipliers with the corresponding binary ones in \cite{Eti2}. In \cite{Eti3}, we presented the best implementation of quaternary adders that is compared to the previously proposed ones\cite{Ebrahimi}\cite{Moaiyeri} \cite{Roosta}. In this paper, we examine quaternary multipliers to check the validity of the previous quote for these important combinational circuits.

The implementation of N-digit adders is easily derived from the implementation of 1-digit adders, with variants to speed-up the carry propagations. Multipliers are most complicated as they involve two steps to multiply N x N digits in the common implementations with minimal propagation delays:
\begin{itemize}
\item Multiply the ith digit by the jth digit for 0<i<N and 0<j<N. It involves $N^2$ 1-digit multipliers
\item Sum the different lines of partial products. Reduction trees such as Wallace or Dada trees \cite{Townsend} are generally used
\end{itemize}
We first present the methodology that is used. Then we present the implementation a 1x1 quaternary digit multiplier. As a N*N multiplier involves both 1-digit multipliers and 1-digit adders, we compare an 8x8 bit multiplier with two  4x4 quaternary digit multipliers. Possible variants of these multipliers are discussed before a final conclusion.

\section{Methodology}
\subsection{Why CNTFET technology?}
This technology uses field-effect transistors that use a single carbon nanotube or an array of carbon nanotubes as the channel material instead of bulk silicon in the traditional MOSFETs. The MOSFET-like CNTFETs having p and n types look the most promising ones. The technology has advantages and drawbacks:
\begin{itemize}
\item CNTFETs have variable threshold voltages (according to the inverse function of the diameter). This is a big advantage compared to CMOS for which different masks are needed to get different threshold voltages. 
\item Among advantages, high electron mobility, high current density, high tranductance can be quoted.
\item Lifetime issues, reliability issues, difficulties in mass production and production costs are quoted as disadvantages.
\item CNTFET technology is far from being a mature one. In 2019, a 16-bit RISC microprocessor has been built with 14,000 CNFET transistors \cite{Hills}. While this is an advance for CNTFET technology, we may observe that the Intel 8086 CPU, which was a 16-bit microprocessor, has been launched in 1978 with 29,000 transistors, more than 40 years ago!
\end{itemize} 
However, as CMOS circuits and CNTFET ones have basically the same circuit styles, CNTFETs can be used to propose a new implementation of quaternary operators and compare it with previous published proposals.

\subsection {Comparing different implementations of quaternary multipliers}
The transistor count is used to compare different implementations of quaternary adders. As comparisons are done by using the same technology and the same operators, the transistor count  is significant as it is very doubtful that more transistors could lead to: 
\begin{itemize}
\item	less interconnects
\item	reduced chip area
\item	reduced power dissipation
\item	reduced propagation delays
\item	Etc.
\end{itemize}

\section{Quaternary circuits}
As previously mentioned, we use the CNTFET technology that as  been used in the most recent papers proposing quaternary adders. In \cite{Eti3}, we have summarized the different techniques to get 4 voltage levels, either with three power supplies or only one power supply, that have been used in papers \cite{Ebrahimi}\cite{Moaiyeri} \cite{Roosta}. In these last two papers that present both versions, the 3 power supplies versions always use less transistors than the 1 power supply one. Version \cite{Roosta} with 3 power supplies is the best direct quaternary implementation. We use this version for the comparison with binary implementations.

\section{Quaternary 1-digit multiplier}
\label{L1}
Table \ref{T1} shows the truth table of a 1-digit quaternary multiplier.
From Table \ref{T1}, we can observe that:
\begin{itemize}
\item When A = 0 then QM = 0 and QC=0
\item When A = 1 then QM = B and QC = 0/0/0/1 for B=0/1/2/3
\item When A = 2 then QM = 0/2/0/2 and QC = 0/0/1/1 for B = 0/1/2/3
\item When A = 3 then QM = 0/3/2/1 and QC= 0/0/1/2 for B = 0/1/2/3
\end{itemize}
\begin{table}
\centering
\caption{Truth table of a quaternary multiplier}
\begin{tabular}{|c|c||c|c|c|c|c||c|c|c|c|}
  \hline
A&Bi&QS&QC&&A&Bi&QM&QC\\
\hline
0&0&0&0&&2&0&0&0\\
0&1&0&0&&2&1&2&0\\
0&2&0&0&&2&2&0&1\\
0&3&0&0&&2&3&2&1\\
1&0&1&0&&3&0&0&0\\
1&1&2&0&&3&1&3&0\\
1&2&3&0&&3&2&2&1\\
1&3&0&1&&3&3&1&2\\
  \hline
\end{tabular}
\label {T1}
\end{table}

\begin{figure}[htbp]
\centerline{\includegraphics  [width =8 cm]{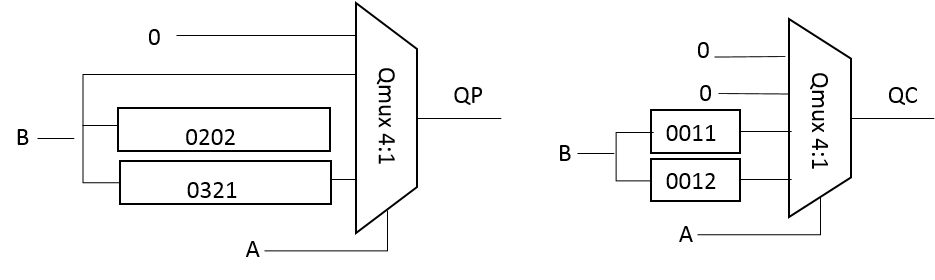}}
\caption{Quaternary multiplier}
\label{QMRoosta}
\end{figure}

\begin{figure}[htbp]
\centerline{\includegraphics  [width = 9 cm]{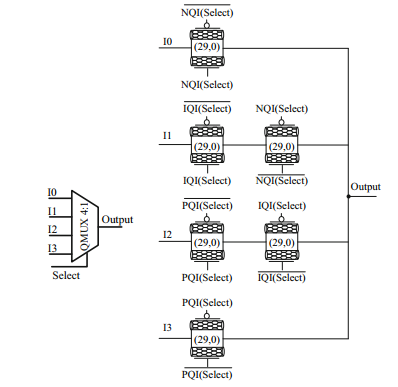}}
\caption{QMUX 4:1 presented in \cite{Roosta}} 
\label{RF3}
\end{figure}

\begin{figure}[htbp]
\centerline{\includegraphics  [width = 9 cm]{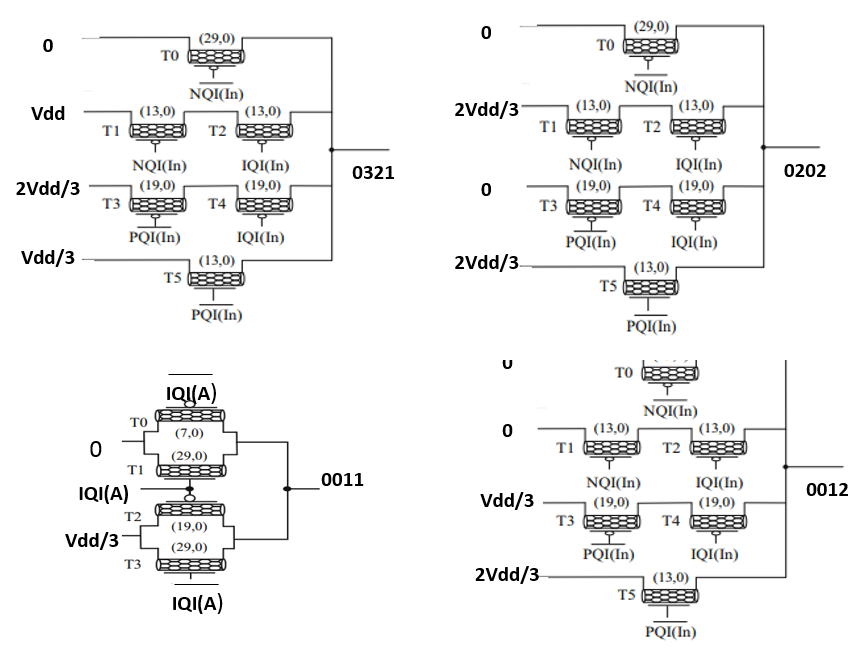}}
\caption{Multiplier subcircuits} 
\label{QMSucc}
\end{figure}

\begin{figure}[htbp]
\centerline{\includegraphics  [width =6 cm]{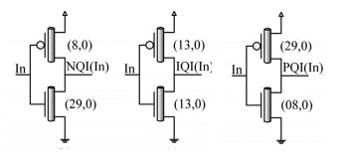}}
\caption{Decoder circuits presented in \cite{Ebrahimi}}
\label{4to2decoder}
\end{figure}

Using the same technique as in \cite{Roosta}, the Product and Carry circuits are shown in Fig. \ref{QMRoosta}. The QMux 4:1 presented in \cite{Roosta} is shown in Fig. \ref{RF3}. The different other circuits used in the multiplier are shown in Fig. \ref{QMSucc}. NQI, IQI and PQI functions correspond to Table \ref{Q2B} in which binary values are 0 and 3. NQI, IQI and PQI outputs are provided by 3 inverters having 3 different threshold levels. Fig. \ref{4to2decoder} shows the corresponding circuits presented in \cite{Ebrahimi}.

\begin{table}
\centering
\caption{Truth table of decoder circuits}
\begin{tabular}{|c||c|c|c|}
  \hline
 IN&NQI&IQI&PQI\\
\hline
 0&3&3&3\\
 1&0&3&3\\
 2&0&0&3\\
 3&0&0&0\\
  \hline
\end{tabular}
\label {Q2B}
\end{table}

The transistor count depends on the layout. Without this layout, it can be evaluated according to two different ways:
\begin{itemize}
\item Count the lower bound of the number of transistors, assuming that fan-out is unlimited and that there is no interconnection issues for the layout. This lower bound is unrealistic. For the multiplier, it means that the same inverter gates (NQI, IQI, PQI, etc) driven by A controls the two Qmux 4:1 and that the same inverter gates driven by B controls the different subcircuits (0202, 0321, 0011, and 0012). 
\item Assume that QP and QC are two different subblocks. There are separate inverter gates controlling the Qmux 4:1 and the subcircuits in the QP and QC blocks.
\end{itemize}

The two corresponding transistor counts are given in Table \ref{T2}. The lower bound in 54 T while 76 T is a more realistic value. The 1x1 bit multiplier is implemented by a AND gates (6T). However, a direct comparison cannot be done, as a NxN digit multiplier uses both 1x1 digit multipliers and 1x1 digit adders. 

\begin{table}
\centering
\caption{Quaternary multiplier transistor count}
\begin{tabular}{|c|c||c|c|c|c|c||c|c||c|c|}
  \hline
Input&Circuits&Min&Subblock\\
\hline
A&NQI,NQI/,IQI,PQI, PQI/&12&24\\
B&NQI,NQI/,IQI,PQI, PQI/&10&20\\
&0202,0321&10&10\\
&0011,0012&10&10\\
&MUX4&12&12\\
&Total&54&76\\
  \hline
\end{tabular}
\label {T2}
\end{table}

\section {A  8 x 8 bit binary multiplier}
\begin{figure}[htbp]
\centerline{\includegraphics  [width = 9 cm]{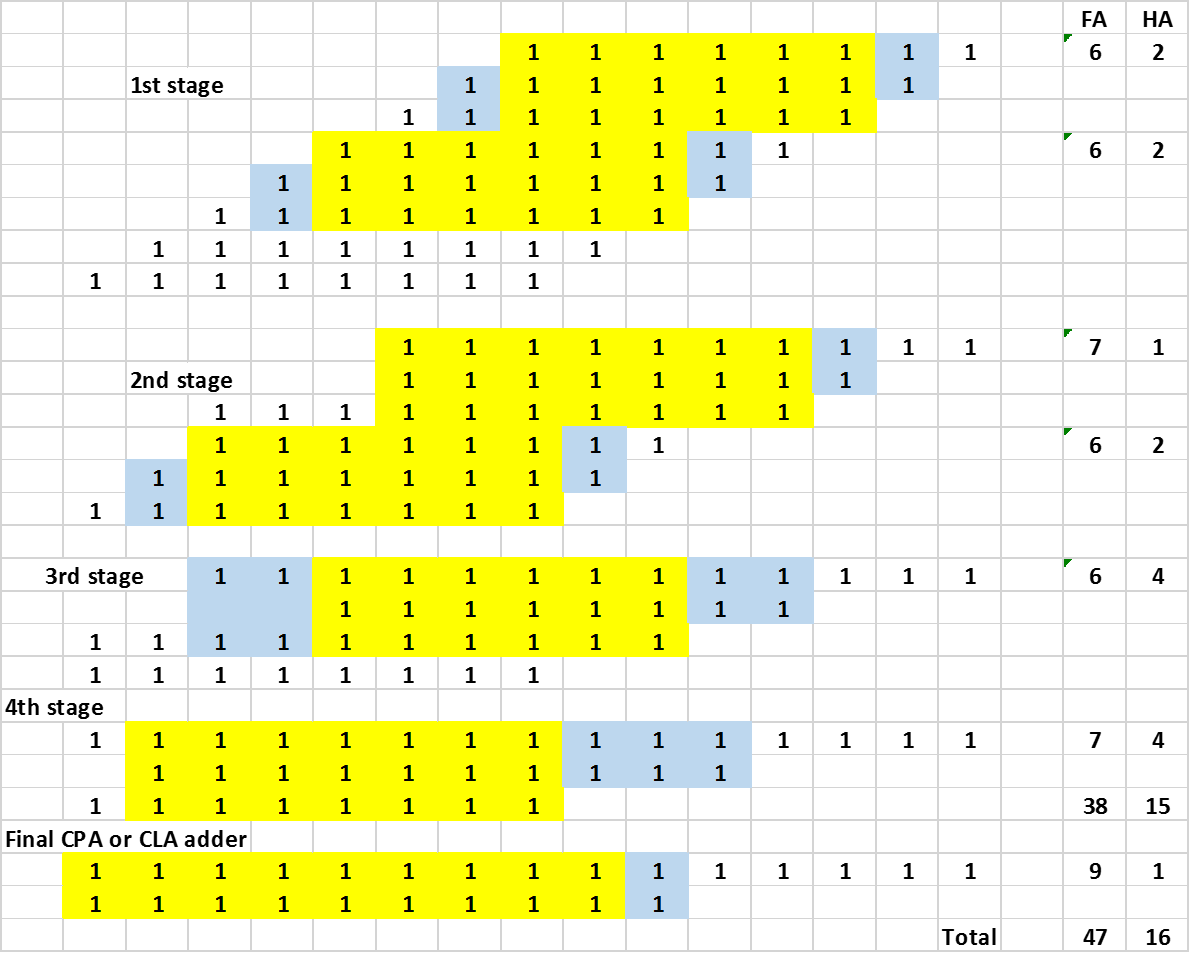}}
\caption{8*8 Wallace Multiplier} 
\label{W88}
\end{figure}

Wallace tree is a typical reduction tree used to implement fast combinational multipliers. Dadda tree is another one.
Fig. \ref{W88} presents this 8*8 multiplier. There are 64 AND gates (1*1 bit multiplier), 38 1-bit adders (FAs) and 15 1-bit half adders (HAs) for the reduction tree. Either a 10-bit Carry Propagate Adder (CPA) or a 10-bit Carry Look-ahead Adder can be used. The CPA would use 9 FAs and 1 HA.

The 1-bit multipliers are implemented with 6 T (Nand + inverter) for a total of 64 x 6 = 384 T. The overall transistor count depends on the implementation of FAs and HAs. A survey was presented in \cite{Anitha}.  The transistor counts range from 28 T for the conventional CMOS design down to 8 T for a scheme using 3T Xor gates. Typical implementations with transmission gates use 14 T or 16 T. All circuits are not equivalent: while conventional CMOS design has maximal noise margins, circuits using transmission gates, or directly connecting inputs either to drain or source of transistors can have reduced noise margins. They can have poor driving capabilitity issues and their switching performance degrades drastically in the cascaded mode
of operation if the suitably designed buffers are not included. For a fair comparison with quaternary circuits, we will consider a 16 T implementation (such as \cite{Bhatta}) and the 28 T of the conventional CMOS design.
The transistor count for the 8 x 8 bit multiplier is given in Table \ref{T3}.
\begin{table}
\centering
\caption{8 x 8 bit multiplier transistor count}
\begin{tabular}{|c|c||c|c|c|c|c||c|c||c|c|}
  \hline
Subcircuit&16 T FAs&28 T FAs\\
\hline
1-bit multiplier&64*6=384 T&64*6=384 T\\
Wallace FAs&38*16=608 T&38*28 = 1064 T\\
Wallace HAs&15*16=240 T& 15*12 = 180 T\\
Final CPAs&10*16 = 160 T&9*28+1*12=264 T \\
Total&1392 T&1892 T\\
  \hline
\end{tabular}
\label {T3}
\end{table}

\section{A 4x4 digit quaternay multiplier with quaternary to binary interfaces}
In this section, we implement a  4x4 digit quaternary multiplier by using a 8x8 bit binary multiplier and quaternary to binary decoders and binary to quaternary encoders.
\subsection{Quaternary to binary interfaces}
These interfaces have already been used in \cite{Eti3}.
\subsubsection{Quaternary to binary decoder}
\begin{figure}[htbp]
\centerline{\includegraphics  [width =6 cm]{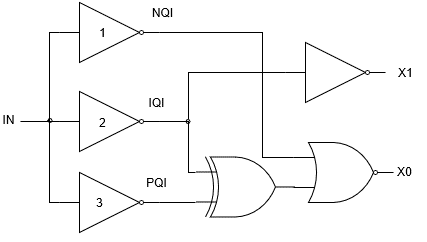}}
\caption{Quaternary to Binary Decoders}
\label{Q2Bdecoder}
\end{figure}

Table \ref{Q2Bconversion} presents the truth table of the quaternary to binary conversion. Binary values are 0 and 3.
The decoder circuit is presented in Fig. \ref{Q2Bdecoder}. The circuitry is the same using 3 or 1 voltage levels. It is based on the inverters 1, 2 and 3 with the different threshold levels (such as the inverters presented in Fig. \ref{4to2decoder}) followed by usual binary gates. The number of transistors depends on the implementation of the XOR gate. It ranges from 16 T when using 4 Nand gates down to 3 T as proposed in \cite{nehru}.% (Fig.\ref{3Txor}). 
An acceptable value is 9 T, which corresponds to the conventional CMOS implementation used in \cite{xor}. This implementation doesn't use pass transistors and has a full swing output.
The overall transistor count for the decoder is then 21 T.

\begin{table}
\centering
\caption{Truth table of decoder circuits}
\begin{tabular}{|c||c|c|c||c|c|}
  \hline
 Q&NQI&IQI&PQI&X1&x0\\
\hline
 0&3&3&3&0&0\\
 1&0&3&3&0&3\\
2&0&0&3&3&0\\
3&0&0&0&3&3\\
  \hline
\end{tabular}
\label {Q2Bconversion}
\end{table}

\subsubsection{Binary to quaternary encoder}
\begin{figure}[htbp]
\centerline{\includegraphics  [width =4 cm]{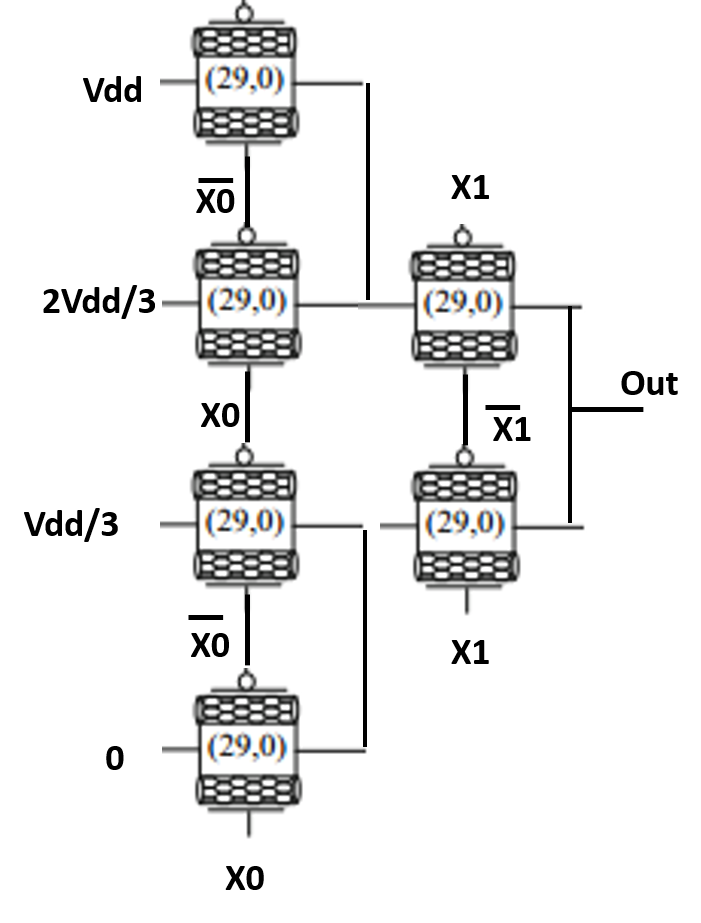}}
\caption{Binary to Quaternary encoder}
\label{B2Qencoder}
\end{figure}

With 3 power supplies, the encoder can be implemented with the mux approach with pass transistors as shown in Fig. \ref{B2Qencoder}. The transistor count including two invertors is 14 T.
\subsubsection{Interface transistor count}
The transistor count for decoding and encoding a quaternary digit is 21 + 14 = 35 T.

\subsection{Quaternary multiplier transistor count}
When using CPAs for the final step of the Wallace tree, the overall transistor count is given in Table \ref{Q2B44M}.

\begin{table}
\centering
\caption{Transistor count for the quaternary multiplier with binary interfaces}
\begin{tabular}{|c||c|c|c||c|c|}
  \hline
 Circuit&16T FA&28T FA\\
\hline
 Interface&35*4 = 140 T&140 T\\
8x8 bit multiplier&1392 T&1892 T\\
Total&1532 T&2032 T\\
  \hline
\end{tabular}
\label {Q2B44M}
\end{table}

\section{Direct implementation of a 4 x 4 digit quaternary multiplier}
We now use the 1-digit multiplier presented in section \ref{L1}. The multiplier has 16 1-digit multipliers. However, as these multipliers generate both a product and a carry output, there are 8 lines of 4 quaternary digits to be reduced by the Wallace tree.
\subsection{1-bit multipliers}
According to Table \ref{T3}, the transistor count is 54 x 16 = 864 T (MIN) or 76 x 16 = 1216 T (Subblock option).

\subsection{Wallace tree and final add}

\begin{figure}[htbp]
\centerline{\includegraphics  [width =8 cm]{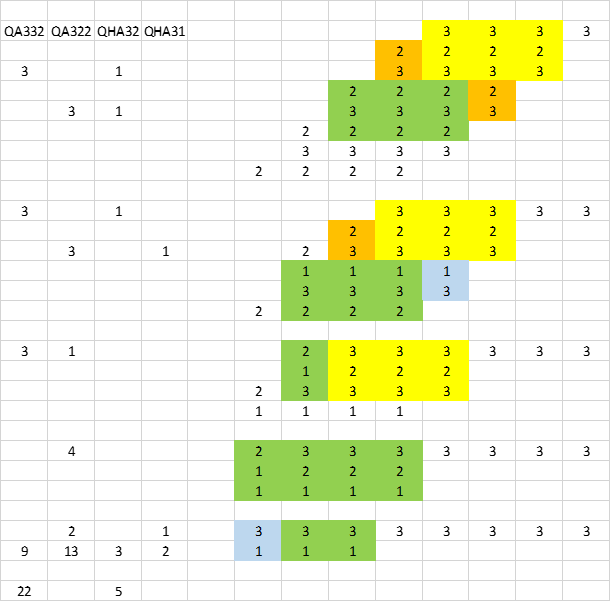}}
\caption{4x4 Wallace Tree}
\label{WQ44}
\end{figure}

The 4x4 Wallace tree with a final stage of CPAs is presented in Fig. \ref{WQ44}. In this figure, 3 corresponds to a quaternary value, 2 to a ternary value and 1 to a binary value. The first stage reduces 8 lines of 4 digits produced by the 16 1-bit multiplier. Lines of 3 correspond to QP quaternary outputs and lines of 2 correspond to QC ternary outputs of the 1-bit multipliers. Using 3 and 2 (max values of quaternary and ternary digits) has the advantage to indicate the different types of adders that must be used. While binary Wallace trees only use binary FAs and HAs, the situation is more complex for quaternary Wallace trees:
\begin{itemize}
\item Q332 adds two quaternary inputs and one ternary input. The ternary input correspond to a QC output of the 1-bit multiplier. As 3+3+2 = 20 (base 4), it means that Q332 generates a quaternary sum and a ternary carry.
\item Q331 adds two quaternary inputs and one binary input. 3+3+1 = 13 (base 4). Q331 generates a quaternary sum and a binary carry. It turns out that Q331 is the quaternary 1-digit adder to be used for N-digit quaternary adders.
\item Q32 adds one quaternary input and one ternary input. 3+2 = 11 (base 4). Q32 generate a quaternary sum and a binary carry. It is a degraded form of the Q33 quaternary half adder.
\item Q31 adds one quaternary input and one binary input to generate a quaternary sum and a binary carry.
\end{itemize}
As shown in Fig. \ref{WQ44}, the Wallace tree with a final CPA has 9 Q332, 13 Q331, 3 Q32 and 2 Q31. There are two options:
\begin{itemize}
\item Using the four different types of adders to minimize the transistor count
\item Only using Q332 and Q32 adders to minimize the different cells and simplify placement and routing of these cells
\end{itemize}

\begin{figure}[htbp]
\centerline{\includegraphics  [width = 9 cm]{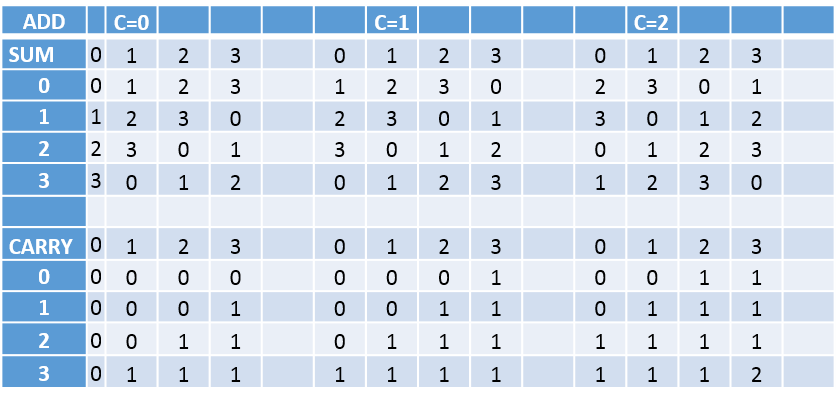}}
\caption{Truth Table of Quaternary Adders.}
\label{TTQADD}
\end{figure}

Fig. \ref{TTQADD} presents the truth table of quaternary adders (sum and carry outputs) when Cin=0/1/2. 

\begin{figure}[htbp]
\centerline{\includegraphics  [width = 9 cm]{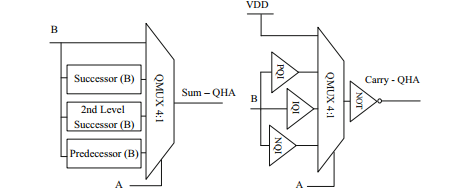}}
\caption{Half adder presented in \cite{Roosta}}
\label{RF1}
\end{figure}

\begin{figure}[htbp]
\centerline{\includegraphics  [width = 9 cm]{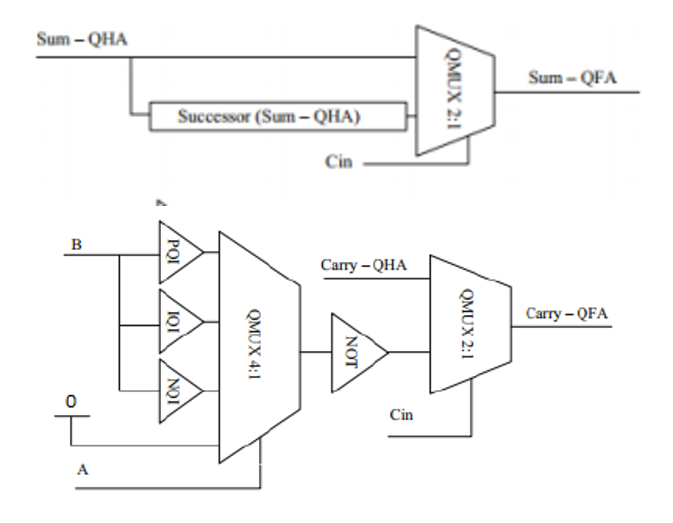}}
\caption{Full adder (Q331)  presented in \cite{Roosta}}
\label{RF2}
\end{figure}

As in \cite{Eti3}, we use the QFA (Q331) adder presented in \cite{Roosta} with three power supplies. The half adder is shown in Fig.\ref{RF1}. It corresponds to the left columns when Cin=0. The full adder is presented in \ref{RF2}. The Sum part of the full adder corresponds to the higher square of the middle colums. It should be noticed that both input and output binary carries use the 0 and 3 values.

\begin{figure}[htbp]
\centerline{\includegraphics  [width = 9 cm]{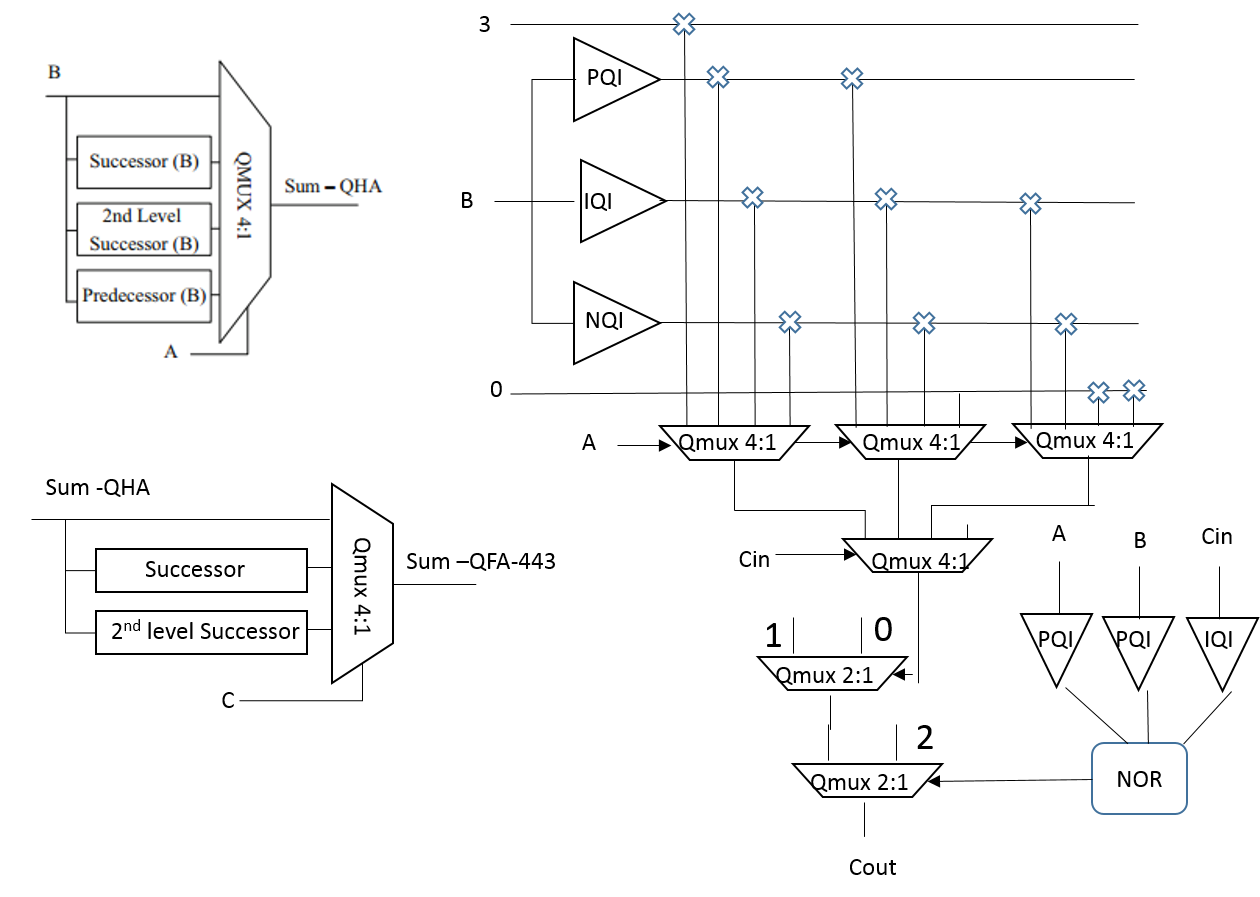}}
\caption{Q332 adder}
\label{Q332}
\end{figure}

The Q332 adder corresponds to the entire truth table. It is shown in Fig. \ref{Q332}. The sum part is easily derived from Fig. \ref{TTQADD}. Now, both input and output carries are ternary. This explain why the Cout computation must generate 0, 1 and 2 levels. 

\begin{figure}[htbp]
\centerline{\includegraphics  [width = 9 cm]{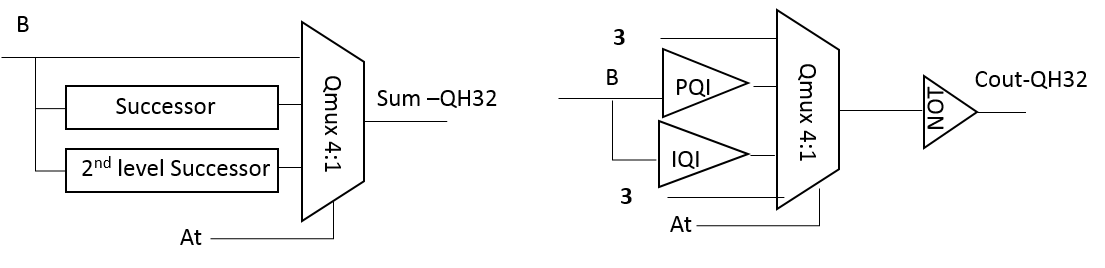}}
\caption{QH32 half adder}
\label{QH32}
\end{figure}

The QH32 half adder is presented in Fig.\ref{QH32}. Cout-QH32 has the binary values 0/3. In the Wallace tree, Cout-QH32 is connected to the Cin input of QH31. It should be noticed that A and B inputs are not symetric: At is a ternary value controling the MUX while B is a quaternary value. 

\begin{figure}[htbp]
\centerline{\includegraphics  [width = 9 cm]{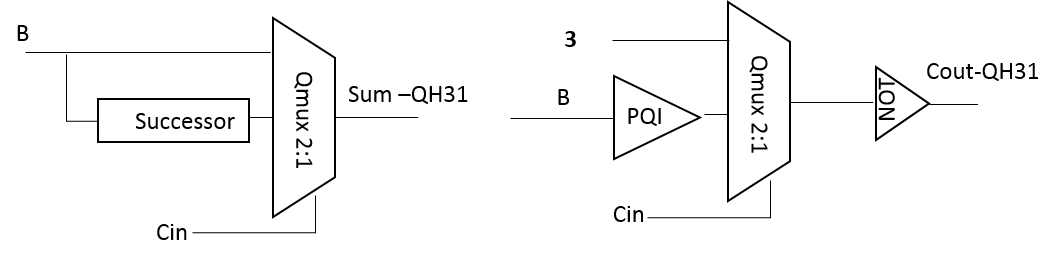}}
\caption{QH31 half adder}
\label{QH31}
\end{figure}

The QH31 half adder is presented in Fig. \ref{QH31} with one quaternary input and one binary input (0/3). Cout is also a 0/3 binary output.

The Q321 adder is similar to Q331 (Fig. \ref{RF2}) except that 0 input in the carry  should be replaced by a Vdd input. So, there is no interest to define a specific Q321 adder.

\begin{table}
\centering
\caption{Quaternary adders transistor count}
\begin{tabular}{|c|c||c|c|c|c|c||c|c||c|c|}
  \hline
Circuits&Min&Subblock\\
\hline
Q331&100&118\\
Q332&154&184\\
QH32&50&54\\
QH31&26&30\\

  \hline
\end{tabular}
\label {TCADD}
\end{table}

The transistor count for these quaternary adders and half adders is given in Table \ref{TCADD}.

The overall transistor count for the 4x4 quaternary digit multiplier is given in Table \ref{TCMUL}

\begin{table}
\centering
\caption{Quaternary multiplier transistor count}
\begin{tabular}{|c|c|c|c|c|c|c|c|c||c|c|}
  \hline
Circuits&Nb&Min&Total &&NB&Subblock&Total\\
\hline
Q331&13&100&1300&&13&118&1534\\
Q332&9&154&1386&&9&184&1656\\
QH32&3&50&150&&3&54&162\\
QH31&2&26&52&&2&30&60\\
\hline
Total&&&2888&&&&3412\\
  \hline
\end{tabular}
\label {TCMUL}
\end{table}

\section{Discussing results}
The different implementations of 4x4 quaternary multipliers compared with 8x8 bit multipliers lead to the following results:
\begin{itemize}
\item A 8x8 bit multiplier has a transistor count ranging from 1392 to 1892 T. The last value corresponds to an implementation using only full CMOS transistor circuitry, without using any pass transistors (Table \ref{T3}).
\item A 4x4 quaternary multiplier implemented with a 8x8 bit multiplier and quaternary encoder and decoder circuit has a transistor count ranging from 1532 to 2032 T, according to the implementation of the binary FAs.
\item A direct implementation of a 4x4 quaternary multiplier has been presented, using the best quaternary approach found in the litterature (\cite{Roosta}). The transistor count ranges from 2888 T (absolute lower bound) to 3412 T (Table \ref{TCMUL})
\end{itemize}
These results lead to the following conclusions:
\begin{itemize}
\item The best implementation of a 4x4 multiplier is obtained by interfacing a 8x8 bit multiplier with quaternary to binary decoder circuits and binary to quaternary encoder circuit. 
\item The most conservative implementation of a 8x8 bit multiplier (1892 T) has far less transistors than the unrealistic lower bound of the transistor count of the direct implementation of 4x4 digit quaternary multiplier (2888 T): 0.65 ratio.
\item As interfacing a binary multiplier with 4-valued interfaces lead to the most efficient implementation, it means that the only gain is to divide by two the number of input and output connections. The overall number of connections is increased: the internal interface connections are added to the binary internal connections. 
\end{itemize}

These results are not surprising. A multiplier consists in 1-digit multipliers and 1-digit adders to reduce the partial products.
\begin{itemize}
\item A quaternary multiplier has 4 times less 1-digit multipliers than the corresponding binary one. It means that a 1-digit multiplier complexity should not be more than 4 times the 1-bit multiplier. While a 1-digit multiplier is a And gate (6 T), the 1-digit quaternary multiplier has a lower bound of 54 T (i.e. x9 ratio). Moreover, the 1-digit multiplier generates a product and a carry.
\item Due to the carry outputs of the 1-digit multiplier, the number of lines in the Wallace tree is exactly the same for the binary and the quaternary cases. When binary lines have N bits, the quaternary lines have N/2 digits. While the 8x8 bit Wallace reduction tree has 38 FAs and 15 HAs, the 4x4 Wallace reduction tree has 9 QA332, 11 QA332, 3 QHA2, and 1 QHA31. The smaller count of quaternary adders cannot compensate the complexity of these adders (Table \ref{TCADD}) versus the complexity of binary FAs and HAs. The results would be similar with the Dadda reduction tree.
\end{itemize}

\section{Concluding remarks}
With the same technique that has been used in the best direct implementation of quaternary adders \cite{Roosta} and the CNTFET circuit styles, we have presented different implementations of a 4x4 digit quaternary multiplier. The best implementation is obtained by using a 8x8 bit multiplier interfaced with quaternary to binary interfaces. It would be the same with any NxN digit multiplier as the issue is with the far greater complexity of 1-digit multipliers and 1-digit adders compared to the complexity of the corresponding binary ones.   
These results are similar with the results already presented in \cite{Eti3} for N-digit adders. 
These results contradict the statement on the advantages of multivalued circuits that is quoted in the introduction. With far more transistors, there is no chance that the quaternary multipliers would have less interconnects, less power dissipation, last chip area than the corresponding binary multipliers.

\end{document}